# Metainference: A Bayesian Inference Method for Heterogeneous Systems

Massimiliano Bonomi[1,*,†], Carlo Camilloni[1,†], Andrea Cavalli[1,2] and Michele Vendruscolo[1,*]

[†] These authors contributed equally to this work

[*] Corresponding author. E-mail: mb2006@cam.ac.uk (M.B.); mv245@cam.ac.uk (M.V.)

[1]Department of Chemistry, University of Cambridge, Cambridge CB2 1EW, UK.

[2]Institute for Research in Biomedicine (IRB), Bellinzona, Switzerland.

## Abstract

Modelling a complex system is almost invariably a challenging task. The incorporation of experimental observations can be used to improve the quality of a model, and thus to obtain better predictions about the behavior of the corresponding system. This approach, however, is affected by a variety of different errors, especially when a system populates simultaneously an ensemble of different states and experimental data are measured as averages over such states. To address this problem we present a Bayesian inference method, called 'metainference', that is able to deal with errors in experimental measurements as well as with experimental measurements averaged over multiple states. To achieve this goal, metainference models a finite sample of the distribution of models using a replica approach, in the spirit of the replica-averaging modelling based on the maximum entropy principle. To illustrate the method we present its application to a heterogeneous model system and to the determination of an ensemble of structures corresponding to the thermal fluctuations of a protein molecule. Metainference thus provides an approach to model complex systems with heterogeneous components and interconverting between different states by taking into account all possible sources of errors.

**Introduction**

The quantitative interpretation of experimental measurements requires the construction of a model of the system under observation. The model usually consists of a description of the system in terms of several parameters, which are determined by requiring the consistency with the experimental measurements themselves as well as with theoretical information, being physical or statistical in nature. This procedure presents several complications. First, experimental data (**Fig. 1A**) are always affected by random and systematic errors (**Fig. 1B, green**), which must be properly accounted for to obtain accurate and precise models. Furthermore, when integrating multiple experimental observations, one must consider that each experiment has a different level of noise so that every element of information is properly weighted according to its reliability. Second, the prediction of experimental observables from the model, which is required to assess the consistency, is often based on an approximate physico-chemical description of a given experiment (the so-called 'forward model') and thus it is intrinsically inaccurate in itself (**Fig. 1B, green**). Third, physical systems in equilibrium conditions often populate a variety of different states whose thermodynamic behaviour can be described by statistical mechanics. In these heterogeneous systems, experimental observations depend from - and thus probe - a population of states (**Fig. 1B, purple**), so that one should determine an ensemble of models rather than a single one (**Fig. 1C**).

Among the theoretical approaches available for model building, two frameworks have emerged as particulary succesfull: Bayesian inference (*1-3*) and the maximum entropy principle (*4*). Bayesian modelling is a rigorous approach to combine prior information on a system with experimental data and to deal with errors in such data (*1-3, 5-8*). It proceeds by constructing a model of noise as a function of one or more unknown uncertainty parameters, which quantify the agreement between predictions and observations and which are inferred along with the model of the system. This method has a long history and it is routinely used in a wide range of applications, including the reconstruction of phyologenetic trees (*9*), the determination of population structures from genotype data (*10*), the interpolation of noisy data (*11*), image reconstruction (*12*), decision theory (*13*), the analysis of microarray data (*14*), and the structure determination of proteins (*15, 16*) and protein complexes (*17*). It has also been extended to deal with mixture of states (*18-21*) by treating the number of states as a parameter to be determined by the procedure. The maximum entropy principle is at the basis of approaches that deal with experimental data averaged over an ensemble of states (*4*) and provides a link between information theory and statistical mechanics. In these methods, an ensemble generated using a prior model is minimally modified by some partial and inaccurate information to match exactly the observed data. In the recently proposed replica-averaging scheme (*22-26*), this result is achieved by modelling an ensemble of replicas of the system using the available information and additional terms that restraint the average values of the predicted data close to the experimental observations. This method has been used to determine ensembles representing the structure and dynamics of proteins (*22-26*).

Each of the two methods described above can deal with some, but not all of the challenges in characterizing complex systems by integrating multiple sources of information (**Fig. 1B**). To overcome all these problems simultaneously, we present the 'metainference' method, a Bayesian inference approach that quantifies the extent to which a prior distribution of models is modified by the introduction of experimental data that are expectation values over a heterogeneous distribution and subject to errors. To achieve this goal, metainference models a finite sample of this distribution, in the spirit of the replica-averaged modelling based on the maximum entropy principle. Notably, our approach reduces to the maximum entropy modelling in the limit of absence of noise in the data, and to standard Bayesian modelling when experimental data are not ensemble averages. This link between Bayesian inference and the maximum entropy principle is not surprising given the connections between these two approaches (*27, 28*). We first benchmark the accuracy of our method on a simple heterogeneous model system, in which synthetic experimental data can be generated with different level of noise as averages over a discrete number of states of the system. We then show its application with nuclear magnetic resonance (NMR) spectroscopy data in the case of the structural fluctuations of the protein ubiquitin in its native state, which we modelled by combining chemical shift with residual dipolar couplings (RDCs).

**Results and Discussions**

Metainference is a Bayesian approach to model a heterogeneous system and all sources of error by considering a set of copies of the system (replicas), which represent a finite sample of the distribution of models, in the spirit of the replica-averaged formulation of the maximum entropy principle (*22-26*). The generation of models by suitable sampling algorithms (typically Monte Carlo or molecular dynamics) is guided by a score given in terms of the negative logarithm of the posterior probability (Materials and Methods):

$$\underbrace{s(\mathbf{X}, \boldsymbol{\sigma})}_{\text{score}} = \underbrace{-\sum_r \log P(X_r, \sigma_r)}_{\substack{r \\ \text{prior}}} + \underbrace{\Delta^2(\mathbf{X})}_{\text{measurements}} \underbrace{\sum_r \frac{1}{2\sigma_r^2}}_{\substack{\text{errors} \\ \sigma_r^2 = (\sigma_r^{SEM})^2 + (\sigma_r^B)^2}}$$

where $\mathbf{X} = [X_r]$ and $\boldsymbol{\sigma} = [\sigma_r]$ are respectively the sets of conformational states and uncertainties, one for each replica. $\sigma_r$ includes all the sources of errors, i.e. the error in representing the ensemble with a finite number of replicas ($\sigma_r^{SEM}$), as well as random, systematic, and forward model errors ($\sigma_r^B$). $P$ is the prior probability which encodes information other than experimental data and $\Delta^2(\mathbf{X})$ is the deviation of the experimental data from the data predicted by the forward model. While this schematic equation, which omits the data likelihood normalization term for the uncertainty parameters, holds for Gaussian errors and a single data point, a more general formulation can be found in the Materials and Methods section, Eqs. 5 and 8.

*Metainference of a heterogeneous model system.* We first illustrate the metainference method for a model system that can populate simultaneously a set of discrete states, i.e. a mixture. In this example, the number of states in the mixture and their population can be varied arbitrarily. We created synthetic data as ensemble averages over these discrete states (**Fig. 2A**) and we added random and systematic noise. We thus introduced a prior information, which provides an approximate description of the system and its distribution of states, and whose accuracy can also be tuned. We then used the reference data to complement the prior information and recover the correct number and populations of the states. We tested the following approaches: metainference (with the Gaussian and outliers noise models in Eqs. 9 and 11, respectively), replica-averaging maximum entropy, and standard Bayesian inference (i.e. Bayesian inference without mixtures).

The accuracy of a given approach was defined as the root mean square deviation (RMSD) of the inferred from the correct populations of the discrete states. We benchmarked the accuracy as a function of the number of data points used, the level of noise in the data, the number of states and replicas, and the accuracy of the prior information. Details of the simulations, generation of data, sampling algorithm, likelihood and model to treat systematic errors and outliers can be found in Supplementary Materials (SM).

*Comparison with the maximum entropy method.* We found that the metainference and the maximum entropy methods perform equally well in absence of noise in the data or in presence of random noise alone (**Fig. 2B,C, grey and orange lines**), as expected, given that maximum entropy is particularly effective in the case of mixtures of states (*22, 23*). The accuracy of the two methods was comparable and, most importantly, increased with the number of data points used (**Fig. 2B,C**). With 20 data points, 128 replicas, and in absence of noise, the accuracy averaged on 300 independent simulations of a 5-state system was equal to 0.4%±0.2% and 0.2%±0.1% for the metainference and maximum entropy approaches, respectively. For reference, the accuracy of the prior information alone was much lower, i.e. 16%. Metainference, however, outperformed the maximum entropy approach in the presence of systematic errors (**Fig. 2B,C, green lines**). The accuracy of metainference increased significantly more rapidly upon adding new information, despite the high level of noise. When using 20 data points, 128 replicas, and 30% outliers ratio, the accuracy averaged on 300 independent simulations of a 5-state system was equal to 2%±2% and 14%±5% for the metainference and maximum entropy approaches, respectively. As systematic errors are ubiquitous, both in the experimental data and in the forward model used to predict the data, this situation reflects more closely a realistic scenario. The ability of metainference to deal at the same time and effectively with conformational averaging and with the presence of systematic errors is the major motivation for introducing this method. This approach can thus leverage the substantial amount of noisy data produced by high-throughput techniques and accurately model conformational ensembles of heterogeneous systems.

*Comparison with standard Bayesian modelling.* In the standard Bayesian approach one assumes the presence of a single state in the sample and estimates its probability or confidence level given experimental data and prior knowledge available. When modelling multiple-states systems with ensemble-averaged data and standard Bayesian modelling, one could be tempted to interpret the probability of each state as its equilibrium population. In doing so, however, one makes a significant error, which grows with the number of data points used, regardless of the level of noise in the data (**Fig. 2D**).

*Role of the prior information.* We tested two priors with different accuracy, with an average population error per state equal to 8% and 16%, respectively. The results suggest that the number of experimental data points required to achieve a given accuracy of the inferred populations depends on the quality of the prior information (**Fig. S1**). The more accurate is the prior, the fewer data points are needed. This is an intuitive, yet important, result. Accurate priors almost invariably require more complex descriptions of the system under study, thus they come at a higher computational costs.

*Scaling with the number of replicas.* As the number of replicas grows, the error in estimating ensemble averages using a finite number of replicas decreases and the overall accuracy of the inferred populations increases (**Fig. S2**), regardless of the level of noise in the data. Furthermore, we verified numerically that, in the absence of random and systematic errors in the data, the intensity of the harmonic restraint, which couples the average of the forward model on the $N$ replicas to the experimental data (Eq. 7), scales as $N^2$ (**Fig. 3**). This test confirms that, in the limit of absence of noise in the data, metainference coincides with the replica-averaging maximum entropy modelling (Materials and Methods).

*Scaling with the number of states.* Metainference is also robust to the number of states populated by the system. We tested our model in the case of 5 and 50 states and determined that the number of data points needed to achieve a given accuracy scales less than linearly with the number of states (**Fig. S3**).

*Outliers model and error marginalization.* As the number of data points and replicas increases, it becomes computationally more and more inconvenient to use one error parameter per replica and data point. In this situation one can assume a unimodal and long-tailed distribution for the errors, peaked around a typical value for a dataset (or experiment type) and replica, and marginalize all the uncertainty parameters of the single data points (Materials and Methods). The accuracy of this marginalized error model was found to be similar to the case in which a single error parameter was used for each data point (**Fig. S4**).

*Analysis of the inferred uncertainties.* We analyzed the distribution of inferred uncertainties $\sigma^B$ in presence of systematic errors (outliers), when using a Gaussian data likelihood with one uncertainty per data point (Eq. 9) and the outliers model with one uncertainty per dataset (Eq. 11). In the former case, metainference was able to automatically detect the data points affected by systematic errors, assign them a higher uncertainty, and thus downweight the associated restraints (**Fig 4A**). In the latter, the inferred typical dataset uncertainty was somewhere in between the uncertainty inferred using the Gaussian likelihood on the data points with no noise and on the outliers (**Fig 4B**). In this specific test (5 states, 20 data points, including 8 outliers, prior accuracy equal to 16%, 128 replicas), both data noise models generated ensemble of comparable accuracy (3%).

**Metainference in integrative structural biology.** We compared the metainference and maximum entropy approaches using NMR experimental data on a classical example in structural biology, the structural fluctuations in the native state of ubiquitin (*22, 29, 30*). A conformational ensemble of ubiquitin was modelled using CA, CB, CO, HA, HN, and NH chemical shifts combined with RDCs collected in a steric medium (*30*) (**Fig. 5A**). The ensemble was validated by multiple criteria (**Table S1**). The stereochemical quality was assessed by PROCHECK (*31*); data not used for modelling, including $^3J_{HNC}$ and $^3J_{HNHA}$ scalar couplings and RDCs collected in other media (*32*), were backcalculated and compared with the experimental data. Exhaustive sampling was achieved by 1 μs long molecular dynamics simulations, performed with GROMACS (*33*) equipped with PLUMED (*34*). As prior information we used the CHARMM22* force field (*35*). Additional details of these simulations can be found in SM.

The quality of the metainference ensemble (**Fig. 5B**) was higher than that of the maximum entropy ensemble, as suggested by the better fit with the data not used in the modelling (**Fig. 5C** and **Table S1**) and by the stereochemical quality (**Table S2**). Data used as restraints were also more accurately reproduced by metainference. One of the major differences between the two approaches is that metainference can deal more effectively with the errors in the chemical shifts calculated on different nuclei. The more inaccurate HN and NH chemical shifts were detected by metainference and thus automatically downweighted in constructing the ensemble (**Fig. 6**).

We also compared the metainference ensemble with an ensemble generated by standard molecular dynamics simulations (MD) and with a high-resolution NMR structure (NMR). The metainference ensemble obtained by combining chemical shifts and RDCs reproduced all the experimental data not used for the modelling better than the MD ensemble and the NMR structure. The only exception were the $^3J_{HNC}$ scalar couplings, which were slightly more accurate in the MD ensemble, and the $^3J_{HNHA}$ scalar couplings, which were better predicted by the NMR structure (**Fig. 5C** and **Table S1**).

The NMR structure, which was determined according to the criterion of maximum parsimony, accurately reproduced most of the available experimental data. Ubiquitin, however, exhibits rich dynamical properties over a wide range of time scales that are averaged in the experimental data (*36*). In particular, a main source of dynamics involves a flip of the backbone of residues D52-G53 coupled with the formation of a hydrogen bond between the side chain of E24 and the backbone of G53. While metainference was able to capture the conformational exchange between these two states, the static representation provided by the NMR structure could not (**Fig. 5B**).

In conclusion, we have presented the metainference approach, which enables building ensemble of models consistent with experimental data when the data are affected by errors and are averaged over a mixtures of states of a system. Since complex systems and experimental data almost invariably exhibit both heterogeneity and errors, we anticipate that our method will find applications across a wide variety of scientific fields, including genomics, proteomics, metabolomics and integrative structural biology.

**Materials and Methods**

The quantitative understanding of a system involves the construction of a model $M$ to represent it. If a system can occupy multiple possible states, one should determine the distribution of models $p(M)$ that specifies in which states the system is found and with which probability. To construct this distribution of models, one should take into account the consistency with the overall knowledge that one has about the system. This includes theoretical knowledge (called the 'prior' information, $I$), and the information acquired from experimental measurements (i.e. the 'data', $D$) (*1*). In Bayesian inference the probability of a model given the information available is known as the *posterior probability* $p(M|D,I)$ of $M$ given $D$ and $I$, and it is given by

$$p(M|D,I) \propto p(D|M,I)p(M|I) \tag{1}$$

where the *likelihood function* $p(D|M,I)$ is the probability of observing $D$ given $M$ and $I$, and the *prior probability* $p(M|I)$ is the probability of $M$ given $I$. To define the likelihood function, one needs a *forward model* $f(M)$ that predicts the data that would be observed for model $M$, and a *noise model* that specifies the distribution of the deviations between the observed and predicted data. In the following we assume that the forward model depends only on the *conformational state* $X$ of the system and that the noise model is defined in terms of unknown parameters $\sigma$ that are part of the model $M = (X,\sigma)$. These parameters quantify the level of noise in the data and they are inferred along with the state $X$ by sampling the posterior distribution. The sampling is usually carried out using computational techniques such as Monte Carlo, molecular dynamics, or combined methods based on Gibbs sampling (*1*).

*Mixture of states.* Experimental data collected under equilibrium conditions are usually the result of ensemble averages over a large number of states. In metainference, the prior information $p(X)$ of state $X$ provides an *a priori* description of the distribution of states. To quantify the fit with the observed data and determine to which extent the prior distribution is modified by the introduction of the data, we need to calculate expectation values of the forward model over the distribution of states. Inspired by the replica-averaged modelling based on the maximum entropy principle (*22-26*), we consider a finite sample of this distribution by modelling simultaneously $N$ replicas of the model $\mathbf{M} = [M_r]$ and we calculate the forward model as an average over the states $\mathbf{X} = [X_r]$

$$f(\mathbf{X}) = \frac{1}{N}\sum_{r=1}^{N} f(X_r) \qquad (2)$$

Typically we have information only about expectation values on the distribution of states $X$, and not on the other parameters of the model, such as $\sigma$. However, we are interested in determining also how the prior distributions of these parameters are modified by the introduction of the experimental data and in doing so we need to treat all the parameters of the model in the same way, i.e. symmetrically. Therefore, we model a finite sample of the distributions of all parameters of the model.

To reduce the computational cost, typically a relatively small number of replicas is used in the modelling. In this situation, the estimate $f(\mathbf{X})$ of the forward model deviates from the average $\tilde{f}$ that would be obtained using an infinite number of replicas. This is an unknown quantity, which we add to the parameters of our model. However, the central limit theorem provides a strong parametric prior since it guarantes that the probability of having a certain value of $\tilde{f}$ given a finite number of states $\mathbf{X}$ is a Gaussian distribution

$$p(\tilde{f} | \mathbf{X}, \sigma^{SEM}) = \frac{1}{\sqrt{2\pi}\sigma^{SEM}} \exp\left[-\frac{\left(\tilde{f} - f(\mathbf{X})\right)^2}{2\left(\sigma^{SEM}\right)^2}\right] \qquad (3)$$

where the standard error of the mean $\sigma^{SEM}$ decreases with the square root of the number of replicas

$$\sigma^{SEM} \propto \frac{1}{\sqrt{N}} \qquad (4)$$

We have recognized so far that in considering a finite sample of our distribution of states we introduce an error in the calculation of expectation values. Therefore, experimental data should be compared to the (unknown) average of the forward model over an infinite number of replicas $\tilde{f}$, which is then related to the average over our finite sample $f(\mathbf{X})$ via the central limit theorem of Eq. 3. From these considerations, we can derive the posterior probability of the ensemble of $N$ replicas representing a finite sample of our distribution of models $\mathbf{M} = (\mathbf{X}, \tilde{f}, \sigma^B, \sigma^{SEM})$. In the case of a single experimental data point $d$, this can be expressed as (see SM)

$$p(\mathbf{X}, \tilde{f}, \sigma^B, \sigma^{SEM} | d, I) \propto \prod_{r=1}^{N} p(d | \tilde{f}_r, \sigma_r^B) \cdot p(\tilde{f}_r | \mathbf{X}, \sigma_r^{SEM}) \cdot p(\sigma_r^B) \cdot p(X_r) \cdot p(\sigma_r^{SEM}) \qquad (5)$$

The data likelihood $p(d|\tilde{f}_r,\sigma_r^B)$ relates the experimental data $d$ to the average of the forward model over an infinite number of replicas, given the uncertainty $\sigma_r^B$. This parameter describes random and systematic errors in the experimental data as well as errors in the forward model. The functional form of $p(d|\tilde{f}_r,\sigma_r^B)$ depends on the nature of the experimental data, and it is typically a Gaussian or log-normal distribution. As noted above, $p(\tilde{f}_r|\mathbf{X},\sigma_r^{SEM})$ is the parametric prior on $\tilde{f}_r$ that relates the (unknown) average $\tilde{f}_r$ to the estimate $f(\mathbf{X})$ computed with a finite number of replicas $N$ via the central limit theorem of Eq. 3, and thus it is always a Gaussian distribution. $p(\sigma_r^{SEM})$ is the prior on the standard error of the mean $\sigma_r^{SEM}$ and encodes Eq. 4, $p(\sigma_r^B)$ is the prior on the uncertainty parameter $\sigma_r^B$, and $p(X_r)$ is the prior on the structure $X_r$.

*Gaussian noise model.* We can further simplify Eq. 5 in the case of Gaussian data likelihood $p(d|\tilde{f}_r,\sigma_r^B)$. In this situation, $\tilde{f}_r$ can be marginalized (see SM) and the posterior probability can be written as

$$p(\mathbf{X},\sigma|d,I) \propto \prod_{r=1}^{N} \frac{1}{\sqrt{2\pi}\sigma_r} \exp\left[-\frac{(d-f(\mathbf{X}))^2}{2\sigma_r^2}\right] p(\sigma_r)p(X_r) \quad (6)$$

where the effective uncertainty $\sigma_r = \sqrt{\left(\sigma_r^{SEM}\right)^2 + \left(\sigma_r^B\right)^2}$ encodes all sources of errors: the statistical error due to the use of a finite number of replicas, experimental and systematic errors, and errors in the forward model. The associated energy function in units of $k_BT$ becomes

$$E = (d-f(\mathbf{X}))^2 \sum_{r=1}^{N} \frac{1}{2\sigma_r^2} + \sum_{r=1}^{N} \left[\log \sigma_r - \log p(\sigma_r) - \log p(X_r)\right] \quad (7)$$

This equation shows how metainference includes different existing modelling methods in limiting cases. In the absence of data and forward model errors ($\sigma_r^B = 0$), our approach reduces to the replica-averaged maximum entropy modelling, in which a harmonic restraint couples the replica-averaged observable to the experimental data. The intensity of the restraint $k = \sum_{r=1}^{N} \frac{1}{\sigma_r^2}$ scales with the number of replicas as $N^2$, i.e. more than linearly, as required by the maximum entropy principle (*24*). We numerically verified this behavior in our heterogeneous model system, in absence of any errors in the data (**Fig. 3**). In presence of errors ($\sigma_r^B \neq 0$), the intensity $k$ scales as $N$ and it is modulated by the data uncertainty $\sigma_r^B$. Finally, in the case in which the experimental data are not ensemble averages ($\sigma_r^{SEM} = 0$), we recover the standard Bayesian modelling.

*Multiple experimental data points.* Eq. 5 can be extended to the case of $N_d$ independent data points $\mathbf{D} = [d_i]$, possibly gathered in different experiments with varying levels of noise (see SM)

$$p(\mathbf{X},\tilde{f},\sigma^B,\sigma^{SEM}|\mathbf{D},I) \propto \prod_{r=1}^{N}\prod_{i=1}^{N_d} p(d_i|\tilde{f}_{r,i},\sigma_{r,i}^B) \cdot p(\tilde{f}_{r,i}|\mathbf{X},\sigma_{r,i}^{SEM}) \cdot p(\sigma_{r,i}^B) \cdot p(\sigma_{r,i}^{SEM}) \cdot \prod_{r=1}^{N} p(X_r) \quad (8)$$

***Outliers model.*** To reduce the number of parameters that need to be sampled in the case of multiple experimental data points, one can model the distribution of the errors around a typical dataset error and marginalize the error parameters for the individual data points. For example, a dataset can be defined as a set of chemical shifts or RDCs on a given nucleus. In this cases, it is reasonable to assume that the level of error of the individual data points in the dataset is homogenoues, except for the presence of few outliers. Let us consider for example the case of Gaussian data noise. In the case of multiple experimental data points, Eq. 6 becomes

$$p(\mathbf{X}, \boldsymbol{\sigma} | \mathbf{D}, I) \propto \prod_{r=1}^{N} p(X_r) \prod_{i=1}^{N_d} \frac{1}{\sqrt{2\pi}\sigma_{r,i}} \exp\left[-\frac{(d_i - f_i(\mathbf{X}))^2}{2\sigma_{r,i}^2}\right] p(\sigma_{r,i}) \quad (9)$$

The prior $p(\sigma_{r,i})$ can be modeled using a unimodal distribution peaked around a typical dataset effective uncertainty $\sigma_{r,0}$ and with a long tail to tolerate outliers data points (*37*)

$$p(\sigma_{r,i}) = \frac{2\sigma_{r,0}}{\sqrt{\pi}\sigma_{r,i}^2} \exp\left(-\frac{\sigma_{r,0}^2}{\sigma_{r,i}^2}\right) \quad (10)$$

where $\sigma_{r,0} = \sqrt{(\sigma^{SEM})^2 + (\sigma_{r,0}^B)^2}$, with $\sigma^{SEM}$ is the the standard error of the mean for all data points in the dataset and replicas and $\sigma_{r,0}^B$ is the typical data uncertainty of the dataset. We can thus marginalize $\sigma_{r,i}$ by integrating over all its possible values, given that all the data uncertainties $\sigma_{r,i}^B$ range from 0 to infinity

$$p(\mathbf{X}, \boldsymbol{\sigma_0} | \mathbf{D}, I) \propto \prod_{r=1}^{N} p(X_r) \cdot \prod_{i=1}^{N_d} \int_{\sigma^{SEM}}^{+\infty} d\sigma_{r,i} \frac{\sqrt{2}\sigma_{r,0}}{\pi\sigma_{r,i}^3} \cdot \exp\left[-\frac{0.5(d_i - f_i(\mathbf{X}))^2 + \sigma_{r,0}^2}{\sigma_{r,i}^2}\right]$$

$$= \prod_{r=1}^{N} p(X_r) \cdot \prod_{i=1}^{N_d} \frac{\sqrt{2}\sigma_{r,0}}{\pi} \cdot \frac{1}{(d_i - f_i(\mathbf{X}))^2 + 2\sigma_{r,0}^2} \cdot \left\{1 - \exp\left[-\frac{0.5(d_i - f_i(\mathbf{X}))^2 + \sigma_{r,0}^2}{(\sigma^{SEM})^2}\right]\right\} \quad (11)$$

After marginalization, we are left with just one parameter $\sigma_{r,0}^B$ per replica that needs to be sampled.

**Acknowledgments**

The authors would like to thank Antonietta Mira for useful Bayesian discussions.

**Figures and Tables**

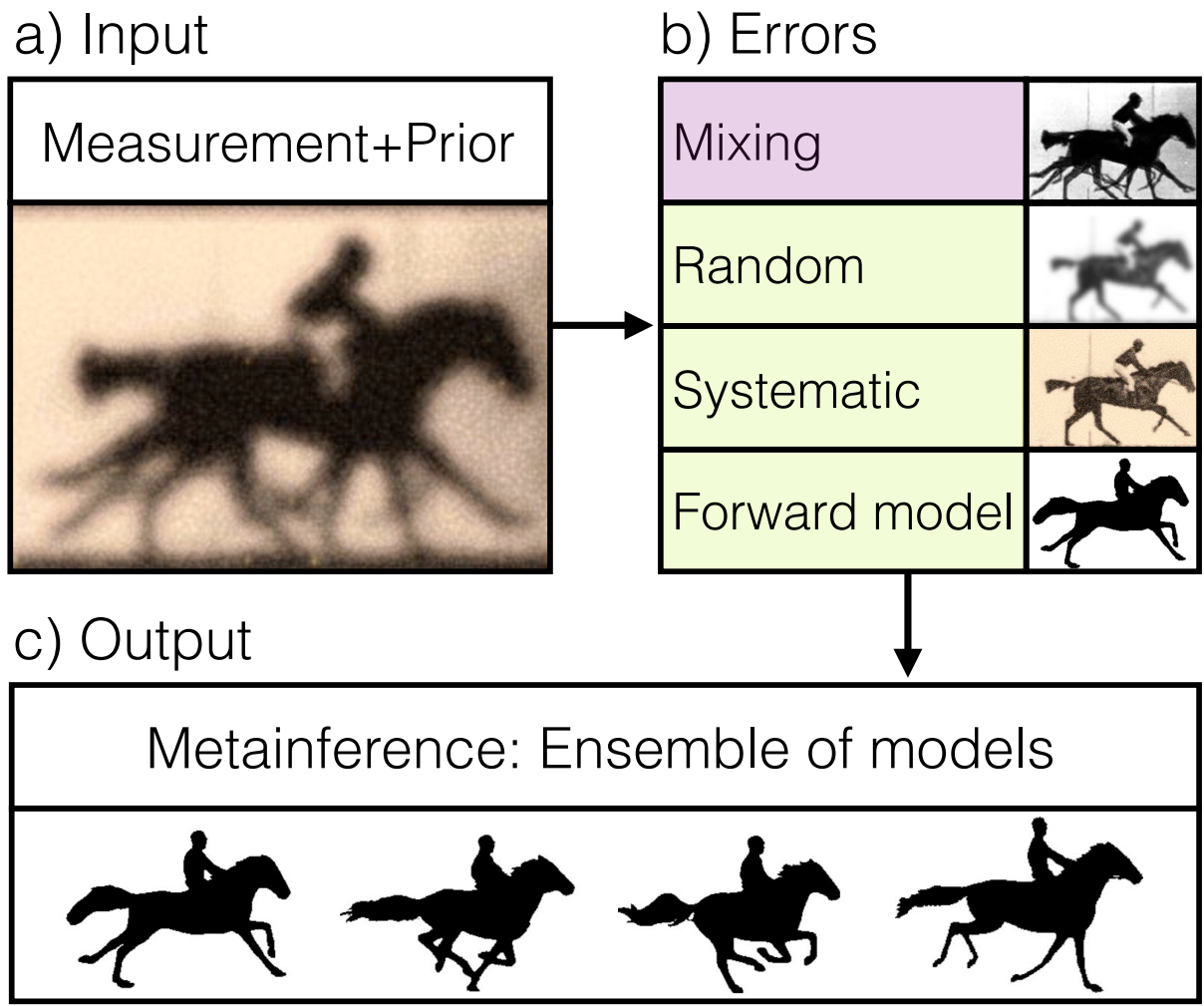

**Figure 1. Schematic illustration of the metainference method.** To generate accurate and precise models from input information (**A**), one must recognize that data from experimental measurements are always affected by random and systematic errors and that the physico-chemical interpretation of an experiment is also inaccurate (**B**, green panels). Moreover, data collected on heterogeneous systems depend on a multitude of states and by their population (**B**, purple panel). Metainference can treat all these sources of error and thus it can properly combine multiple experimental data with prior knowledge of a system to produce ensembles of models consistent with the input information (**C**).

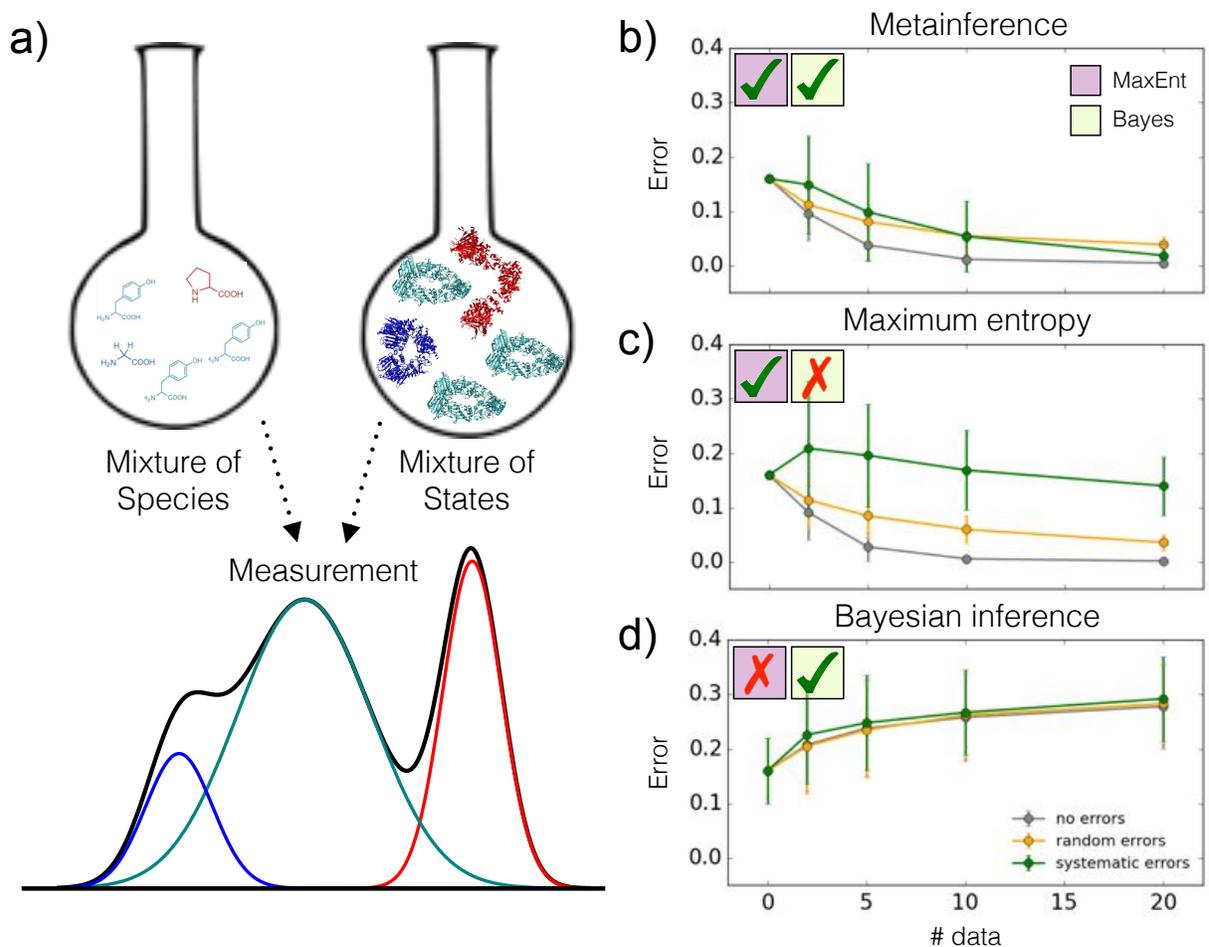

**Figure 2. Metainference of a model heterogeneous system.** Equilibrium measurements on mixtures of different species or states do not reflect a single specie or conformation, but are instead averaged over the whole ensemble (**A**). We describe such a scenario using a model heterogeneous system composed of multiple discrete states on which we tested metainference (**B**), the maximum entropy approach (**C**), and standard Bayesian modelling (**D**), using synthetic data. We assess the accuracy of these methods in determining the populations of the states as a function of the number of data points used and the level of noise in the data. Among these approaches metainference is the only one that can deal with both heterogeneity and errors in the data; the maximum entropy approach can treat only the former, while standard Bayesian modelling only the latter.

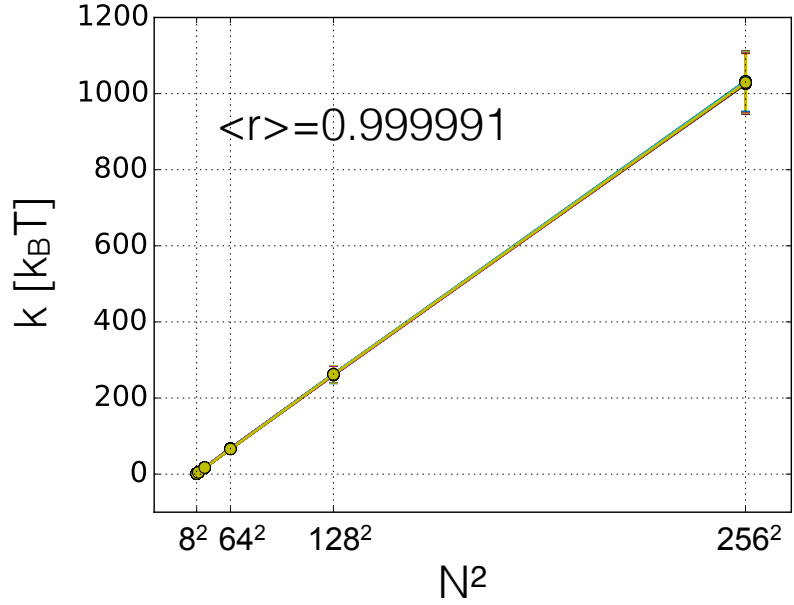

**Figure 3. Scaling of the matainference harmonic restraint intensity in absence of noise in the data.** We verified numerically that in absence of noise in the data, and with a Gaussian noise model, the intensity of the metainference harmonic restraint $k = \sum_{r=1}^{N} \frac{1}{\sigma_r^2}$, which couples the average of the forward model over the $N$ replicas to the experimental data point (Eq. 7), scales as $N^2$. This test was carried out in the model system at 5 discrete states, with 20 data points and the prior with accuracy equal to 16%. For each of the 20 data points, we report the average restraint intensity over the entire Monte Carlo simulation and its standard deviation, when using 8, 16, 32, 64, 128, and 256 replicas. The average Pearson's correlation coefficient on the 20 data points is $0.999991 \pm 3 \cdot 10^{-6}$, showing that metainference coincides with the replica-averaging maximum entropy modelling in the limit of absence of noise in the data.

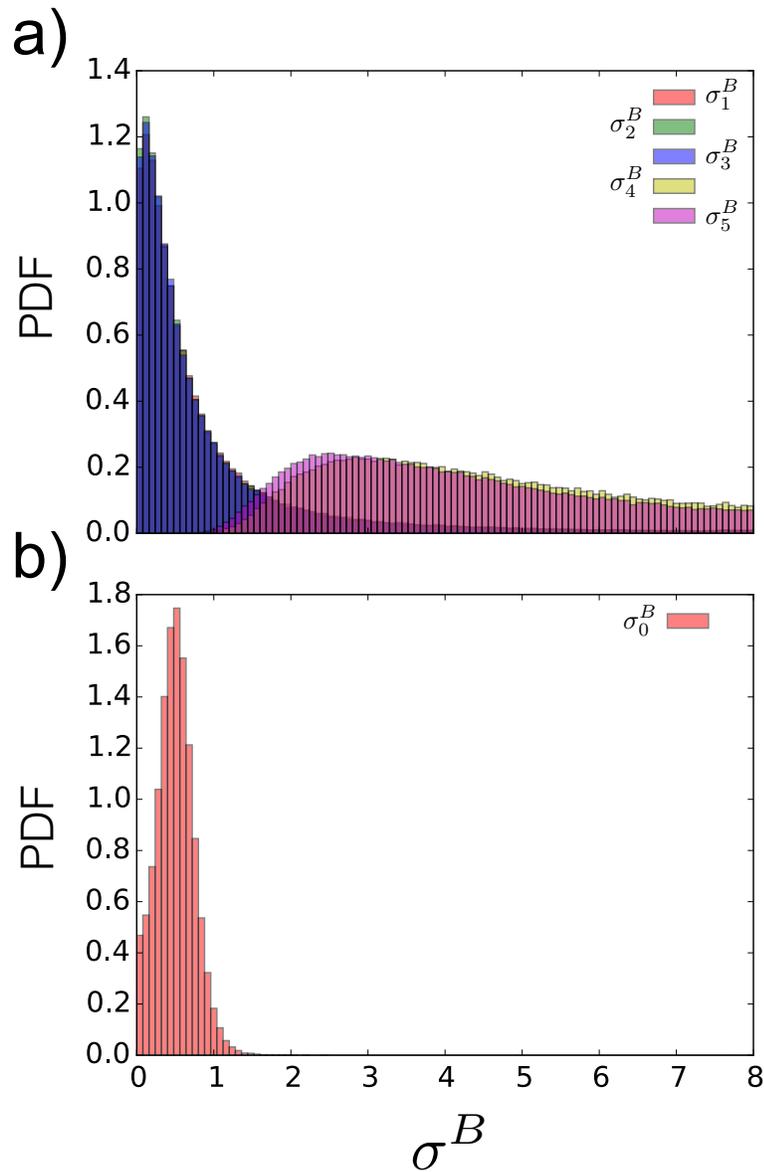

**Figure 4. Analysis of the inferred uncertainties.** Distributions of inferred uncertainties in presence of systematic errors, (**A**) using a Gaussian data likelihood with one uncertainty per data point and (**B**) the outliers model with one uncertainty per dataset. This test was carried out in the model system at 5 discrete states, with 20 data points (of which 8 were outliers), 128 replicas, and the prior with accuracy equal to 16%. For the Gaussian noise model, we report the distributions of 3 representative points not affected by noise ($\sigma^B_{1-3}$) and of 2 affected by systematic errors ($\sigma^B_4$ and $\sigma^B_5$). For the outliers model, we report the distribution of the typical dataset uncertainty ($\sigma^B_0$).

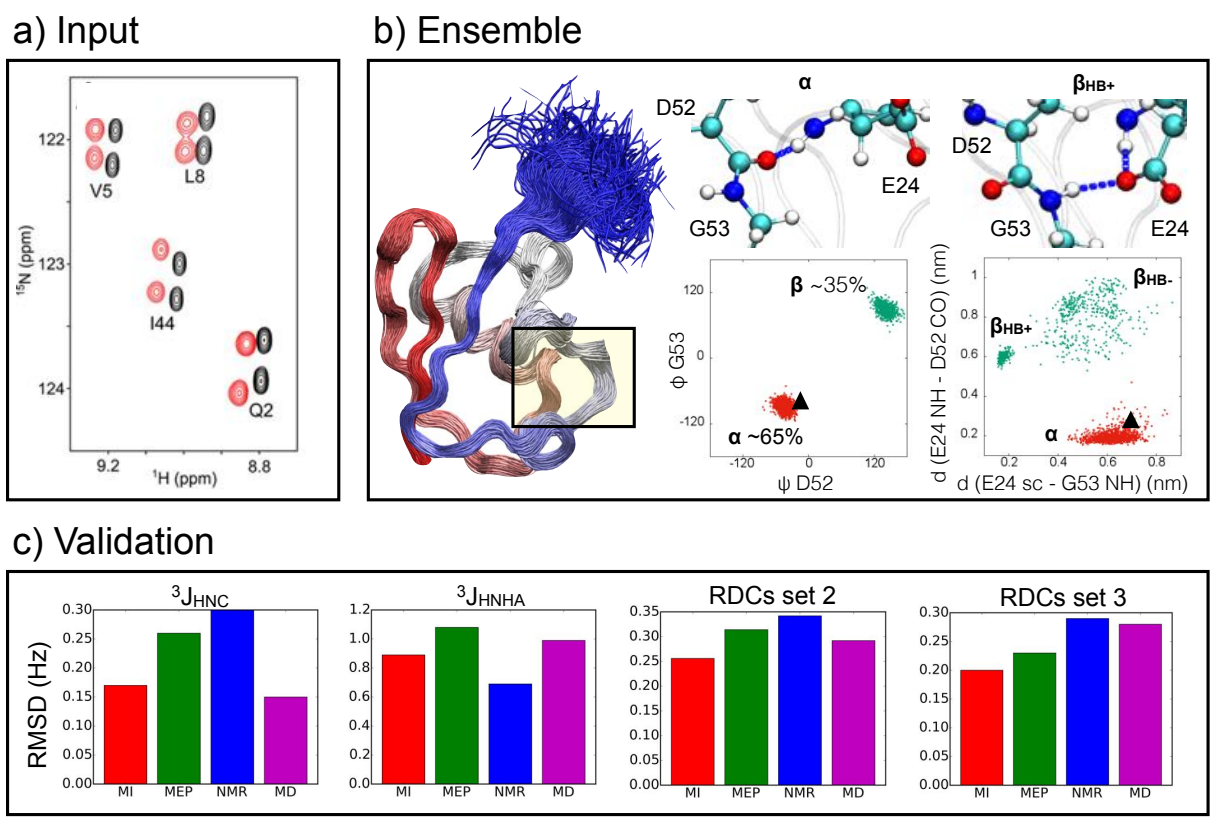

**Figure 5. Example of the application of metainference in integrative structural biology.** (**A**) Comparison of the metainference and maximum entropy approaches by modelling the structural fluctuations of the protein ubiquitin in its native state using NMR chemical shifts and RDC data. (**B**) The metainference ensemble supports the finding (*36*) that a major source of dynamics involves a flip of the backbone of residues D52-G53 (**B**, left scatter plot), which interconverts between an α state with a 65% population and a β state with a 35% population. This flip is coupled with the formation of a hydrogen bond between the side-chain of E24 and the backbone of G53 (**B**, right scatter plot); the state in which the hydrogen bond is present ($\beta_{HB+}$) is populated 30% of the time, and the state in which the hydrogen bond is absent ($\beta_{HB-}$) is populated 5% of the time. By contrast the NMR structure (PDB code 1D3Z) provides a static picture of ubiquitin in this region in which the α state is the only populated one (black triangle). (**C**) Validation of the metainference (red) and maximum entropy (green) ensembles, along with the NMR structure (blue) and the MD ensemble (purple), by the back-calculation of experimental data not used in the modelling: $^3J_{HNC}$ and $^3J_{HNHA}$ scalar couplings and two independent sets of RDCs (RDCs sets 2 and 3).

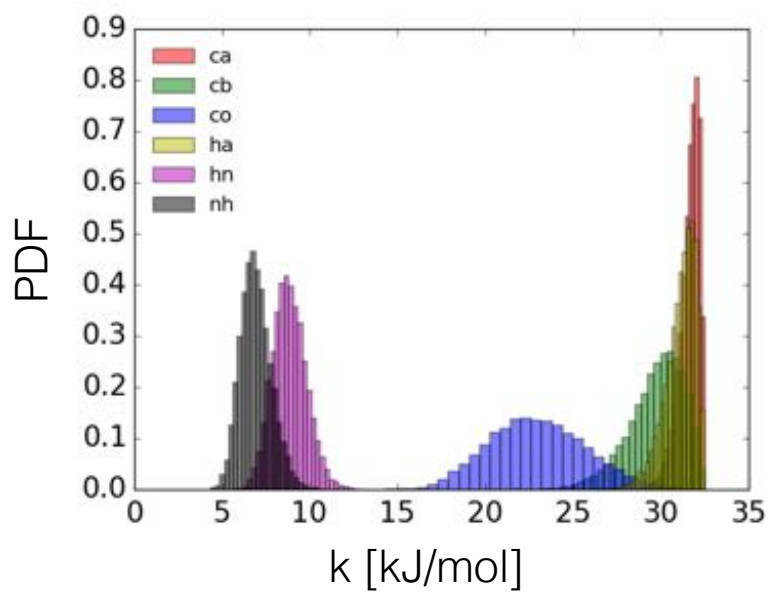

**Figure 6. Distributions of restraint intensities for different chemical shifts of ubiquitin.** When combining data from different experiments, metainference automatically determines the weight of each piece of information. In the case of ubiquitin, the NH and HN chemical shifts were determined as the noisiest data and thus downweighted in the construction of the ensemble of models. At this stage, it is impossible to determine whether these two specific datasets have a higher level of random or systematic noise, or whether instead the CAMSHIFT predictor (*38*) is less accurate for these specific nuclei.

## Supplementary Materials

## Derivation of the basic metainference equations

*1) The metainference posterior in the case of a single experimental data point.* Here we derive Eq. 5 of the main text, which is the general metainference equation in the case of a single experimental data point $d$. As discussed in the main text (Materials and Methods), we are interested in determining how the prior distribution of models (including structural states and other parameters) is affected by the introduction of experimental information. Since experimental data in equilibrium conditions are the result of ensemble averages over a distribution of states, we model a finite sample of the distribution of models, which we refer to as the set of $N$ replicas of the system. These include: the coordinates of the system $\mathbf{X} = [X_r]$, the averages of the forward model over an infinite number of replicas $\tilde{f} = [\tilde{f}_r]$, the uncertainty parameters that describes random and systematic errors in the experimental data as well as errors in the forward model $\boldsymbol{\sigma}^B = [\sigma_r^B]$, the standard errors of the mean $\boldsymbol{\sigma}^{SEM} = [\sigma_r^{SEM}]$.
The metainference posterior probability is thus

$$p(\mathbf{X}, \tilde{f}, \boldsymbol{\sigma}^B, \boldsymbol{\sigma}^{SEM} \mid d, I) \tag{S1}$$

We first recognize that $\mathbf{X}$ and $\boldsymbol{\sigma}^{SEM}$ do not dependent from the data $d$. Therefore

$$p(\mathbf{X}, \tilde{f}, \boldsymbol{\sigma}^B, \boldsymbol{\sigma}^{SEM} \mid d, I) = p(\tilde{f}, \boldsymbol{\sigma}^B \mid d, I) \cdot p(\mathbf{X}) \cdot p(\boldsymbol{\sigma}^{SEM}) \tag{S2}$$

At this point, we should take into account that each set $\tilde{f} = [\tilde{f}_r]$, $\boldsymbol{\sigma}^B = [\sigma_r^B]$, and $\boldsymbol{\sigma}^{SEM} = [\sigma_r^{SEM}]$ is composed of independent variables, and that the configurations $\mathbf{X} = [X_r]$ are a priori independent. Given these considerations, we can write from Eq. S2

$$p(\mathbf{X}, \tilde{f}, \boldsymbol{\sigma}^B, \boldsymbol{\sigma}^{SEM} \mid d, I) = \prod_{r=1}^{N} p(\tilde{f}_r, \sigma_r^B \mid d, I) \cdot p(X_r) \cdot p(\sigma_r^{SEM}) \tag{S3}$$

By applying Bayes theorem to $p(\tilde{f}_r, \sigma_r^B \mid d, I)$ we can thus derive Eq. 5 of the main text

$$p(\mathbf{X}, \tilde{f}, \boldsymbol{\sigma}^B, \boldsymbol{\sigma}^{SEM} \mid d, I) \propto \prod_{r=1}^{N} p(d \mid \tilde{f}_r, \sigma_r^B) \cdot p(\tilde{f}_r \mid \mathbf{X}, \sigma_r^{SEM}) \cdot p(\sigma_r^B) \cdot p(X_r) \cdot p(\sigma_r^{SEM}) \tag{S4}$$

*2) Gaussian data noise and marginalization.* We can further simply Eq. S4 in the case of Gaussian data likelihood

$$p(d \mid \tilde{f}_r, \sigma_r^B) = \frac{1}{\sqrt{2\pi}\sigma_r^B} \cdot \exp\left[-\frac{(d - \tilde{f}_r)^2}{2(\sigma_r^B)^2}\right] \tag{S5}$$

In this case, we can write

$$p(d | \tilde{f}_r, \sigma_r^B) \cdot p(\tilde{f}_r | \mathbf{X}, \sigma_r^{SEM}) = \frac{1}{\sqrt{2\pi}\sigma_r^B} \cdot \exp\left[-\frac{(d-\tilde{f}_r)^2}{2(\sigma_r^B)^2}\right] \cdot \frac{1}{\sqrt{2\pi}\sigma_r^{SEM}} \cdot \exp\left[-\frac{(\tilde{f}_r - f(\mathbf{X}))^2}{2(\sigma_r^{SEM})^2}\right] \quad (S6)$$

The product of the two Gaussian probability density functions (PDFs) is a scaled Gaussian PDF

$$p(d | \tilde{f}_r, \sigma_r^B) \cdot p(\tilde{f}_r | \mathbf{X}, \sigma_r^{SEM}) = \frac{S}{\sqrt{2\pi}\bar{\sigma}} \cdot \exp\left[-\frac{(\tilde{f}_r - \bar{f})^2}{2\bar{\sigma}^2}\right] \quad (S7)$$

where

$$\bar{\sigma} = \sqrt{\frac{(\sigma_r^B)^2 \cdot (\sigma_r^{SEM})^2}{(\sigma_r^B)^2 + (\sigma_r^{SEM})^2}} \quad \text{and} \quad \bar{f} = \frac{d \cdot (\sigma_r^{SEM})^2 + f(\mathbf{X}) \cdot (\sigma_r^B)^2}{(\sigma_r^B)^2 + (\sigma_r^{SEM})^2} \quad (S8)$$

The scaling factor is itself a Gaussian PDF

$$S = \frac{1}{\sqrt{2\pi\left((\sigma_r^{SEM})^2 + (\sigma_r^B)^2\right)}} \cdot \exp\left[-\frac{(d - f(\mathbf{X}))^2}{2\left((\sigma_r^{SEM})^2 + (\sigma_r^B)^2\right)}\right] \quad (S9)$$

Since typically we are not interested in determining $\tilde{f}_r$, we can marginalize it as

$$\int p(d | \tilde{f}_r, \sigma_r^B) \cdot p(\tilde{f}_r | \mathbf{X}, \sigma_r^{SEM}) \cdot d\tilde{f}_r = S = \frac{1}{\sqrt{2\pi}\sigma_r} \cdot \exp\left[-\frac{(d - f(\mathbf{X}))^2}{2\sigma_r^2}\right] \quad (S10)$$

where the effective uncertainty parameters $\sigma_r = \sqrt{(\sigma_r^{SEM})^2 + (\sigma_r^B)^2}$ encodes all sources of error. If we incorporate Eq. S10 into Eq. S4 we obtain the marginalized version of Eq. 5 that holds for Gaussian data noise (Eq. 6 in the main text).

*3) The metainference posterior in the case of multiple independent data points.* We now extend Eq. S4 to the case of $N_d$ independent data points $\mathbf{D} = [d_i]$. We thus introduce one $\tilde{f}$, $\sigma_{r,i}^B$, and $\sigma_{r,i}^{SEM}$ per data point $i$ and replica $r$. In this case $\tilde{\mathbf{f}} = [[\tilde{f}_{r,i}]]$, $\boldsymbol{\sigma}^B = [[\sigma_{r,i}^B]]$, and $\boldsymbol{\sigma}^{SEM} = [[\sigma_{r,i}^{SEM}]]$.

Since $\mathbf{X}$ and $\boldsymbol{\sigma}^{SEM}$ do not dependent from the data $\mathbf{D}$, the posterior can be written as

$$p(\mathbf{X}, \tilde{\mathbf{f}}, \boldsymbol{\sigma}^B, \boldsymbol{\sigma}^{SEM} | \mathbf{D}, I) = p(\tilde{\mathbf{f}}, \boldsymbol{\sigma}^B | \mathbf{D}, I) \cdot p(\mathbf{X}) \cdot p(\boldsymbol{\sigma}^{SEM}) \quad (S11)$$

Each set $\tilde{f} = [[\tilde{f}_{r,i}]]$, $\sigma^B = [[\sigma^B_{r,i}]]$, and $\sigma^{SEM} = [[\sigma^{SEM}_{r,i}]]$ is composed of independent variables, and the configurations $\mathbf{X} = [X_r]$ are a priori independent. Therefore we can write

$$p(\mathbf{X}, \tilde{f}, \sigma^B, \sigma^{SEM} | \mathbf{D}, I) = \prod_{r=1}^{N} \prod_{i=1}^{N_d} p(\tilde{f}_{r,i}, \sigma^B_{r,i} | \mathbf{D}, I) \cdot p(\sigma^{SEM}_{r,i}) \cdot \prod_{r=1}^{N} p(X_r) \quad \text{(S12)}$$

By applying Bayes theorem to the data likelihood $p(\tilde{f}_{r,i}, \sigma^B_{r,i} | \mathbf{D}, I)$, we can write

$$p(\mathbf{X}, \tilde{f}, \sigma^B, \sigma^{SEM} | \mathbf{D}, I) \propto \prod_{r=1}^{N} \prod_{i=1}^{N_d} p(\mathbf{D} | \tilde{f}_{r,i}, \sigma^B_{r,i}) \cdot p(\tilde{f}_{r,i} | \mathbf{X}, \sigma^{SEM}_{r,i}) \cdot p(\sigma^B_{r,i}) \cdot p(\sigma^{SEM}_{r,i}) \cdot \prod_{r=1}^{N} p(X_r) \quad \text{(S13)}$$

We now use the fact that the multiple data points are independent to factorize the data likelihood

$$p(\mathbf{D} | \tilde{f}_{r,i}, \sigma^B_{r,i}) = \prod_{j=1}^{N_d} p(d_j | \tilde{f}_{r,i}, \sigma^B_{r,i}) \quad \text{(S14)}$$

and since the data point $d_j$ depends only on $\tilde{f}_{r,j}$ and $\sigma^B_{r,j}$, we can write

$$p(\mathbf{D} | \tilde{f}_{r,i}, \sigma^B_{r,i}) = p(d_i | \tilde{f}_{r,i}, \sigma^B_{r,i}) \cdot \prod_{j=1, j\neq i}^{N_d} p(d_j) \propto p(d_i | \tilde{f}_{r,i}, \sigma^B_{r,i}) \quad \text{(S15)}$$

By inserting Eq. S15 into Eq. S13 we obtain the metainference equation for the case of multiple independent data points (Eq. 8 in the main text)

$$p(\mathbf{X}, \tilde{f}, \sigma^B, \sigma^{SEM} | \mathbf{D}, I) \propto \prod_{r=1}^{N} \prod_{i=1}^{N_d} p(d_i | \tilde{f}_{r,i}, \sigma^B_{r,i}) \cdot p(\tilde{f}_{r,i} | \mathbf{X}, \sigma^{SEM}_{r,i}) \cdot p(\sigma^B_{r,i}) \cdot p(\sigma^{SEM}_{r,i}) \cdot \prod_{r=1}^{N} p(X_r) \quad \text{(S16)}$$

**Details of the model system simulations.**

To assess the accuracy of the different modelling approaches considered in this work, we studied a model system characterized by multiple discrete states, for which the number of states $N_S$ and their population $[w^0]$ can be varied arbitrarily. This system captures some of the complexity of real mixtures of different species and/or conformations in which equilibrium measurements mix contributions from all states. A simulation of this model system consists of 4 steps.

*1) Generation of states and synthetic experimental data.* For each state, we randomly extracted its population $w^0_k$ and $N_d$ real numbers $d_{i,k}$ in the range from 1.0 to 10.0. These numbers are the pure experimental data points for each state and they will be used as forward model in the next step. The pure *observed* data points are a mixture on all states, $d_i = \sum_{k=1}^{N_S} w^0_k \cdot d_{i,k}$. We introduced two types of noise to the pure observed data points to mimic the presence of random and systematic errors. Random errors were modeled with a Gaussian noise with standard deviation equal to 0.5, while systematic errors were modeled by adding a random offset in the range from 3.0 to 5.0 to 30% of the data points. We modeled systems of 5 states using 2, 5, 10, and 20 data points and systems of 50 states using 20, 50, 100, and 200 data points. For both model sizes, we genereted 4 datasets: (i) without errors, (ii) with only random errors, (iii) with only systematic errors, and (iv) with both random and systematic errors.

*2) Scoring.* In metainference, the total energy of the system is defined as

$$E = -k_B T \cdot \log p(\mathbf{X}, \sigma | \mathbf{D}, I) \tag{S17}$$

where we used a Gaussian noise with one uncertainty parameter $\sigma_{r,i}$ per replica and data point (Eq. 9) or an outliers model with one uncertainty parameter per dataset (Eq. 11). In both cases, we used a Jeffrey's prior $p(\sigma) = 1/\sigma$ for the uncertainty of each data point or for the typical dataset uncertainty. $\sigma^{SEM}$ was kept fixed and equal to $\tilde{\sigma}^{SEM} / \sqrt{N}$, with $\tilde{\sigma}^{SEM} = 5.66$. For the standard Bayesian modelling, $\sigma^{SEM}$ was set to zero. For the replica-averaged approach, we introduced harmonic restraints to couple forward model predictions to the observed data points. The intensity of the harmonic restraints was set to $k = N^2 \cdot k_1$, with $k_1 = 0.03$. We used the same prior information for the metainference, standard Bayesian modelling, and replica-averaged approaches. We randomly perturbed the exact populations $w_k^o$ to obtain approximate weights $w_k$ for each state and thus we defined the energy associated to the prior information as $E_k = -k_B T \cdot \log w_k$. To study the effect of the prior accuracy, we created high and low accuracy priors, with an average population error per state equal to 0.08 and 0.16, respectively.

*3) Sampling.* We simulated $N$ copies of the system to benchmark the metainference and replica-averaged approaches, and a single replica for standard Bayesian modelling. In the former case, we used 8, 16, 32, 64, and 128 replicas. The following unknown variables were sampled by Monte Carlo: a discrete index that determines which state of the system is occupied and the data uncertainty parameters for metainference and standard Bayesian modelling. The data uncertainty parameters were sampled in the range 0.00001-200, by proposing random moves at most equal to 10.0. $k_B T$ was set to 1.0. A total of 50,000 Monte Carlo steps were carried out in each simulation.

*4) Analysis.* During each Monte Carlo simulation we accumulated the histogram of the discrete variable that indicates which state of the system is istantaneously populated. From this histogram, we calculated the population of each state $\tilde{w}_k$ determined from prior information and experimental data. We defined as accuracy the root mean squared deviation of $[\tilde{w}_k]$ from the exact populations $[w_k^o]$. For each approach to test and choice of parameters (number of data points, level of noise in the data, and number of replicas), we run 300 indipendent simulations with random reference state populations and data points. The reported accuracy is averaged over the 300 simulations.

**Details of the molecular dynamics simulations of ubiquitin**

Classical all-atom molecular dynamics simulations of ubiquitin were perfomed using GROMACS (*33*) together with PLUMED (*34*). The X-ray structure 1UBQ (*29*) has been used as starting point in the simulations, using the CHARMM22* force field (*35*), in a cubic box of 6.3 nm of side with 7800 TIP3P water molecules (*39*). A time step of 2 fs was used together with LINCS constraints (*40*). The van der Waals and Coulomb interactions were cut-off at 0.9 nm, while long-range electrostatic effects was treated with the particle mesh Ewald method. All simulations were carried out in the canonical ensemble by keeping the volume fixed and by thermosetting the system at 300 K with the Bussi-Donadio-Parrinello thermostat (*41*). A 1 ms molecular dynamics simulation was perfomed as a reference sampling of the *a priori* information of the CHARMM22* force field.

Maximum entropy replica-averaged simulations and metainference replica-averaged simulations were perfomed using backbone chemical shifts (bmr17760) and residual dipolar couplings measured in a liquid-crystallin phase (N-H, C$\alpha$-H$\alpha$, C$\alpha$-C′, C′-N, C′-H and C$\alpha$-Cb bonds) as structural restraints modelled with CamShifts and the exact $\vartheta$-method, respectively. Maximum entropy and metainference simulations were performed using 8 replicas, in all cases for a total simulation time of 1 ms, consistently with the reference sampling. A Gaussian noise model with one error parameter per nucleus was used in the metainference approach, along with a Jeffrey's prior on each error parameter.

From the resulting ensembles we back-calculated chemical shifts using SPARTA+ (*42*), RDCs measured in a large number of conditions (*32*) using PALES in the SVD approximation (*43*) using only data for residues 1 to 70 to obtain the alignment tensor. Scalar couplings across hydrogen bonds have been calculated as $^{h3}JNC = (-357\ Hz)exp(-3.2r_{HO}/Å)cos^2\theta$ where $\theta$ represents the H…O=C angle (*44*), while H-H$\alpha$ scalar coupling have been calculated using the Karplus equation with previously reported parameters (*45*). In addition the presence of distorted geometries have been tested with PROCHECK (*31*). The ensembles have been also compared with the 1UBQ X-ray (*29*) and the 1D3Z NMR (*30*) structures.

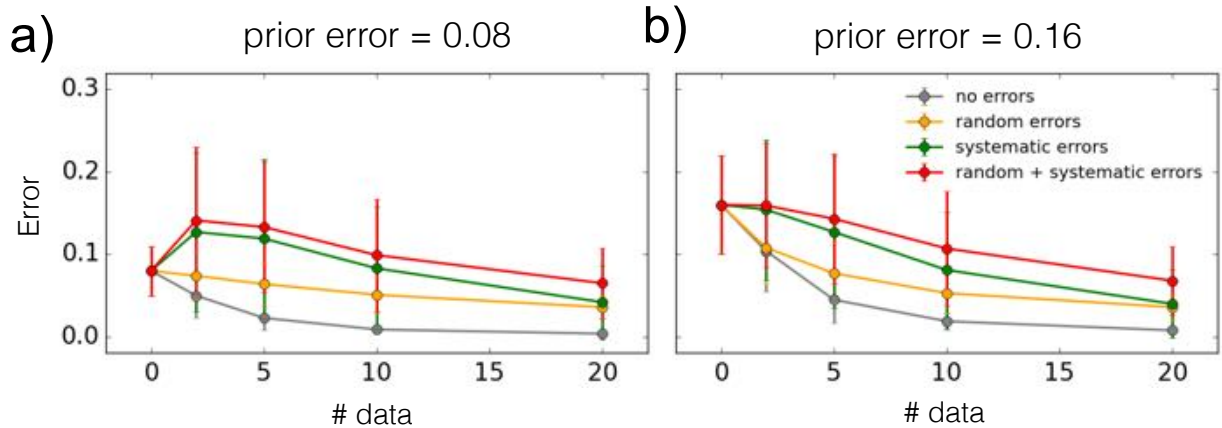

**Figure S1. Effect of prior accuracy on the error of the metainference method.** Metainference error as a function of the number of data points and for varying levels of noise in the data in the case of prior with average error in the state populations equal to 0.08 (**A**) and 0.16 (**B**). The quality of the prior information influences the number of data points required to achieve a given accuracy of the inferred state populations. The more accurate is the prior, the fewer data points are needed. These simulations were carried out on a 5-state model, using 128 replicas.

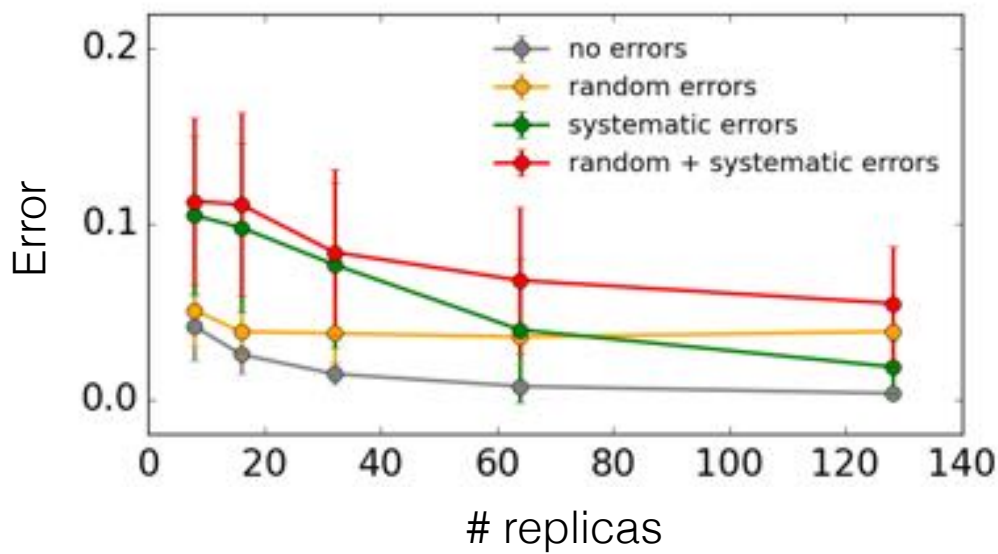

**Figure S2. Scaling of metainference error with number of replicas for varying level of noise in the data.** As the number of replicas increases, the statistical error in calculating ensemble averages with a finite number of replicas converges to zero, and the overall accuracy of metainference increases. These simulations were carried out on a 5-state model, using 20 data points and the prior with average error equal to 0.16.

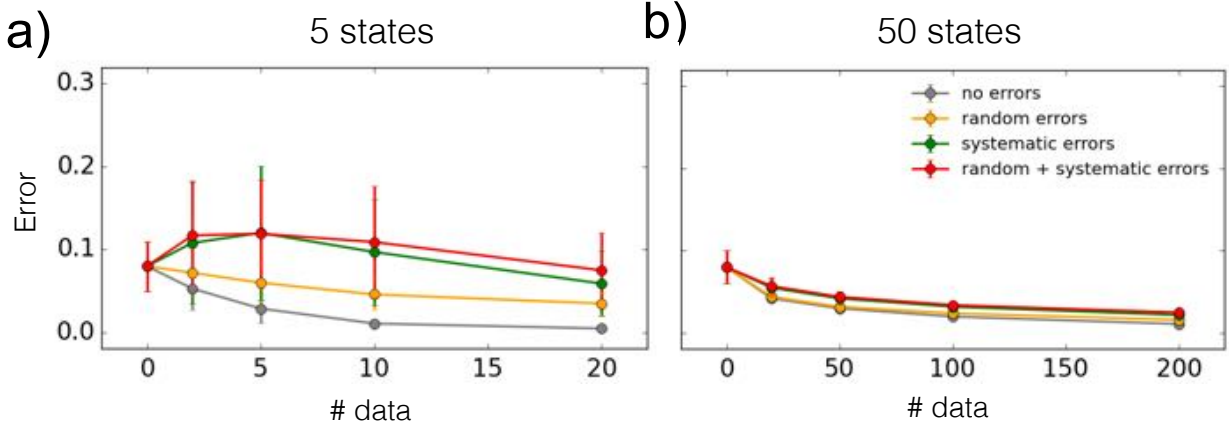

**Figure S3. Scaling of metainference error with number of states.** The metainference error as a function of the number of data points and for varying levels of noise in the data for a system composed of 5 (**A**) and 50 (**B**) states. These simulations were carried using 64 replicas and the prior with average error equal to 0.08.

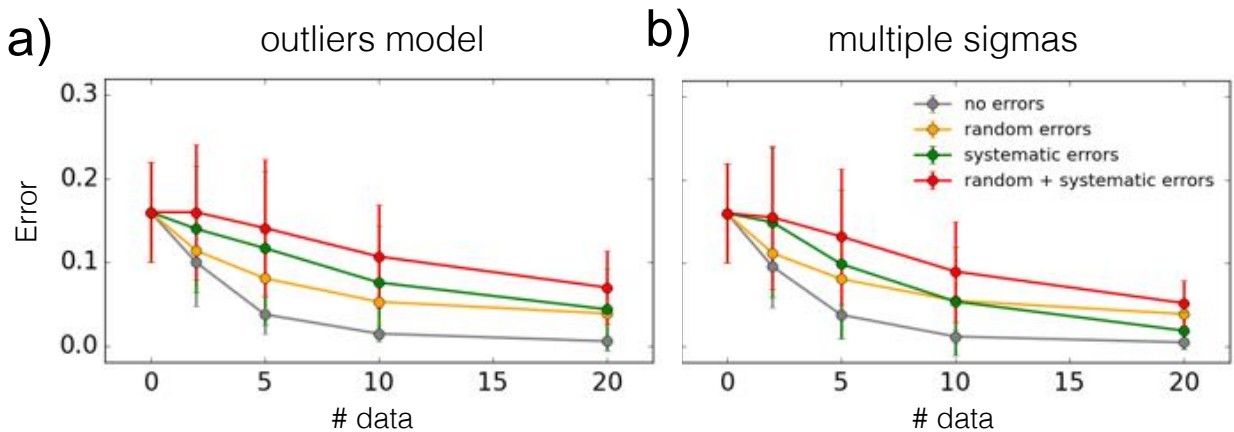

**Figure S4. Accuracy of the outliers model.** Metainference error as a function of the number of data points and for varying levels of noise in the data with an outlier model for the errors that uses a single error parameter per dataset (**A**) and with one error parameter per data point (**B**). These simulations were carried out on a 5-state model, using 128 replicas and the prior with average error equal to 0.16.

| Score | Maximum entropy | Metainference | NMR | MD | X-ray |
|---|---|---|---|---|---|
| | | | Modelling | | |
| **Chemical shifts** | | | | | |
| CA | 0.76 | 0.72 | 0.63 | 0.90 | 0.71 |
| CB | 0.89 | 0.93 | 0.89 | 1.12 | 0.94 |
| CO | 0.80 | 0.81 | 0.75 | 0.93 | 0.80 |
| HA | 0.13 | 0.13 | 0.21 | 0.23 | 0.17 |
| HN | 0.34 | 0.39 | 0.40 | 0.44 | 0.39 |
| NH | 2.51 | 2.32 | 1.86 | 2.77 | 2.03 |
| **RDC set 1** | | | | | |
| NH | 0.16 | 0.15 | 0.19 | 0.27 | 0.21 |
| CAC | 0.13 | 0.13 | 0.27 | 0.23 | 0.31 |
| CAHA | 0.15 | 0.15 | 0.13 | 0.23 | 0.28 |
| CN | 0.15 | 0.14 | 0.23 | 0.24 | 0.21 |
| CH | 0.52 | 0.18 | 0.29 | 0.31 | 0.32 |
| | | | Validation | | |
| $^3J_{HNC}$ | | | | | |
| RMSD | 0.26 | 0.17 | 0.30 | 0.15 | 0.22 |
| $^3J_{HNHA}$ | | | | | |
| RMSD | 1.08 | 0.89 | 0.69 | 0.99 | 0.89 |
| **RDC set 2** | | | | | |
| NH(36) | 0.23 | 0.20 | 0.29 | 0.28 | 0.29 |
| **RDC set 3** | | | | | |
| NH | 0.32 | 0.24 | 0.24 | 0.24 | 0.29 |
| CAC | 0.27 | 0.22 | 0.28 | 0.24 | 0.32 |
| CAHA | 0.37 | 0.33 | 0.40 | 0.32 | 0.42 |
| CN | 0.27 | 0.23 | 0.28 | 0.32 | 0.33 |
| CH | 0.34 | 0.26 | 0.51 | 0.34 | 0.47 |

**Table S1. Comparison of the quality of the ensembles obtained using different modelling approaches in the case of the native state of the protein ubiquitin.** Maximum entropy and metainference indicate the ensembles generated in this work using 8 replicas and chemical shifts combined with RDCs. NMR, MD and X-ray indicate a structure determined using high-resolution NMR methods (PDB code 1D3Z (*30*)), an ensemble determined by standard molecular dynamics simulations, and a X-ray structure (1UBQ) (*29*), respectively. In the upper part of the Table ("Modelling") we report the fit with the data used in the modelling, in the lower part ("Validation") the fit with independent data not used in the modelling.

| Score | Maximum Entropy | Metainference | NMR | MD | X-ray |
|---|---|---|---|---|---|
| **Procheck** | | | | | |
| RAMA | 1.6 | 1.2 | 1.0 | 1.0 | 1.0 |
| HBGEO | 2.3 | 1.9 | 1.4 | 2.1 | 1.7 |
| CHI-1 | 1.4 | 1.4 | 1.0 | 1.3 | 2.0 |
| CHI-2 | 1.0 | 1.0 | 1.0 | 1.0 | 1.4 |
| OMEGA | 2.5 | 2.0 | 1.0 | 2.0 | 1.0 |

**Table S2. Comparison of the stereochemical quality of the ensembles or single models generated by the approaches defined in Table S1.** The quality was assessed with PROCHECK (*31*).